# Observation of two collapsed phases in CaRbFe$_4$As$_4$

Ryan L. Stillwell,[1] Xiangfeng Wang,[2] Limin Wang,[2] Daniel J. Campbell,[2] Johnpierre Paglione,[2,3] Samuel T. Weir,[1] Yogesh K. Vohra,[4] Jason R. Jeffries [1]

[1]*Lawrence Livermore National Laboratory, Livermore, California 94550, USA*

[2]*Center for Nanophysics and Advanced Materials, Department of Physics, University of Maryland, College Park, Maryland, 20742, USA*

[3]*Canadian Institute for Advanced Research, Toronto, Ontario M5G 1Z8, Canada*

[4]*Department of Physics, University of Alabama at Birmingham, Birmingham, Alabama 35294, USA*

**Abstract**

We report the observation of the pressure-induced, fully-collapsed tetragonal phase of CaRbFe$_4$As$_4$ for P~ 22 GPa via high-pressure x-ray diffraction and magnetotransport measurements. The x-ray measurements, along with resistivity measurements, show that there is an initial half-collapsed tetragonal phase for 6 < P < 22 GPa, in which superconductivity is continuously suppressed from $T_c$= 35K at P= 3.1 GPa to $T_c$ <2K for P ≥17.2 GPa, as well as signs of the fully-collapsed tetragonal phase near P=22 GPa. Density functional calculations suggest that both of these transitions are driven by increased As-As bonding, first across the Ca layer, and then at the second transition, across the Rb layer. Although electrical resistivity measurements in the fully-collapsed tetragonal phase do not show superconductivity, there is a change in the slope of both the Hall coefficient and the longitudinal resistance near 22 GPa, suggesting a strong correlation between the electronic and lattice degrees of freedom in this new iron-based superconductor.

## I. Introduction

It has been just over a decade since the discovery of iron-based superconductors opened up a new avenue of investigation into the interplay of superconductivity and magnetism. [1-3] The progress of these materials moved quickly by application of pressure and doping to maximize $T_c$, but then hit a ceiling and rested near 55K.[4-6] Some of the foundational insights into these materials have come from the $Ae$Fe$_2$As$_2$ (122) class (where $Ae$=Ca, Sr, Ba), which evince a wide range of phases such as antiferromagnetism, ferromagnetism, spin-density waves and unconventional superconductivity through chemical doping and the applications of uniaxial and hydrostatic pressure.[7-17]

Recently, a new structural type of FeSCs having the formula $AeA$Fe$_4$As$_4$ was discovered, opening up a new chapter of research in these "1144" FeSCs [18-28]. Similarly to the 122s, these materials form in the ThCr$_2$Si$_2$ crystal structure with layers of Fe$_2$As$_2$ tetrahedral nets, separated by alternating layers of alkaline ($A$) or alkaline-earth $Ae$ metal ions, and similarly to the 122s (with the exception of (Sr,Ba)Fe$_2$As$_2$[8, 29, 30]), superconductivity is observed at ambient pressure, though in nearly all of the 1144 compounds the critical temperature is significantly increased to ~35K without chemical doping or the application of high pressure [18, 19, 31, 32].

Many of the 122s also showed a correlation between a change in magnetism and/or superconductivity with a structural phase transition into a collapsed tetragonal phase with the application of pressure[10, 33-35]. A systematic study of the $A$Fe$_2$As$_2$ series showed that the critical pressure for the cT phase transition was proportional to the cation radius[36] and, in turn, the As-As spacing[10, 37-39].

In this paper we report our study of CaRbFe$_4$As$_4$ (CaRb1144) via high-pressure x-ray diffraction and high-pressure electronic transport measurements up to 35 GPa, to investigate the effects of bond length on structural phase stability, electronic correlations and magnetic fluctuations. We find that there is a half-collapsed tetragonal (hcT) phase transition near 6 GPa and a fully-collapsed tetragonal (fcT) transition near 22 GPa. In both of these collapses we see the expansion along the *a*-axis and contraction along the *c*-axis, though both theory and experiment show a small change in *c* compared to the expansion of the *a*-axis. Similar to predictions and other experimental results on CaK1144, there is a concomitant suppression of T$_c$ with increasing pressure at the hcT transition[20, 21]. This most likely signals the destruction of bulk superconductivity at 5 GPa, similar to the findings of Kaluarachchi et al.[20] and the predictions of Borisov et al..[22] Filamentary superconductivity is then suppressed completely near the fcT transition. Our magnetotransport measurements show a discontinuous decrease in resistivity at both the hcT and fcT transitions, with the Hall coefficient showing a change in the dominant carrier type near the hcT transition. We confirm our experimental results with density-functional theory (DFT) calculations which predict the changes in the *a* and *c* axes as a function of pressure and also show that the Fermi surface is dramatically modified as CaRb1144 is driven through the hcT and fcT transitions.

## II.    Experimental Details

Single crystals of CaRb1144 were prepared in excess FeAs flux. Elemental Ca and Rb were combined with prereacted FeAs in an alumina crucible in a 0.8:1.2:10 ratio, with another crucible placed on top to ensure the volatile material did not escape. The two crucibles were then sealed in a capped stainless steel tube in a glove box of inert nitrogen atmosphere and placed in a furnace. The growth was heated to 650 C in one hour, held at that temperature for three hours, heated to 1180 C over two hours and held there for five, cooled over two hours to 1030 C then more slowly (30 hours) to 900 C, at which point the furnace was turned off and cooled naturally to room temperature. Crystals were then extracted out of the flux at room temperature, allowing for the possible inclusion of 122 phases that are known to form at lower temperatures[40]. The stoichiometry of the resulting crystals was verified using energy dispersive spectroscopy and found that the Fe deficiency relative to As was less than 5%.

Angle-dispersive x-ray diffraction measurements under pressure were performed using beamline 16 BM-D (HPCAT) of the Advanced Photon Source at Argonne National Laboratory. A gas-membrane-driven diamond anvil cell (DAC) with 300μm diamond culets was used to generate pressures up to 30 GPa. The sample was powdered and loaded into the 130-μm diameter sample chamber that was drilled out of a rhenium gasket preindented to 40 μm, along with a Cu powder (3-6$\mu$m, Alfa Aesar) as the pressure calibrant and neon, precompressed to 40,000 psi, as the pressure-transmitting medium. The sample was illuminated with a 0.4246 Å (29.2 keV) monochromatic x-ray beam and angular dispersive diffraction patterns were collected with a Mar345 image plate using an exposure time of 240 s. The two dimensional diffraction images were integrated using Fit2D[41], and pressure- dependent lattice parameters were extracted by indexing the positions of the Bragg reflections using the EXPGUI/GSAS package[42, 43]. To determine the pressure, we used the equation of state of the copper, with fitting parameters for the bulk modulus $B_0$=133 GPa and its pressure derivative $B_0$'=5.01 from Dewaele et al.[44]

Ambient pressure resistivity measurements were performed in a Quantum Design Physical Property Measurement System (PPMS) in zero magnetic field from 300K to 2K, using the standard four-probe configuration of the AC Resistivity option. For electrical transport studies under pressure we used an eight-probe designer DAC [45, 46] with 280μm diameter culets, steatite as a pressure-transmitting medium and ruby as the pressure calibrant. [47, 48] A MP35N metal gasket was preindented to an initial thickness of 40 μm, and a 130 μm hole was drilled in the center of the indentation for the sample chamber using an electric discharge machine. A small crystal of CaRbFe$_4$As$_4$ with dimensions of 70μm x 70μm on a side and 15-μm thick was placed onto the designer anvil to ensure electrical contact with the tungsten leads exposed on the face of the designer diamond culet. Pressure was measured at room temperature on two separate ruby spheres within the sample chamber in order to estimate pressure distribution across the chamber. Based on previous studies using this type of DAC the error in the pressure at low temperatures was estimated to be 5%[46]. All of the error bars for pressure are based on this correction. Temperature was measured using a calibrated Cernox thermometer affixed to the outside of the DAC. Electrical transport measurements were made as a function of temperature and magnetic field using the AC Transport option in the Quantum Design PPMS.

In order to theoretically investigate the pressure-dependent electronic and structural properties of CaRbFe$_4$As$_4$, we employed the same model as in Ref. [20], in which the twisted long-range magnetic order was considered to preserve the tetragonal symmetry of the system, given that the paramagnetic fluctuation played an important role in the structural features of iron pnictides[22]. As shown in Fig. S1 of the supplement[49], the Fe moments are lying in the plane and twisted around the c-axis with the spin either along the a-axis or b-axis, while those from the nearest layer are in the opposite direction. Then the lattice parameters and internal atomic positions were fully relaxed using the Vienna *ab initio* Simulation Package (VASP) [50-52], with the projector-augmented wave basis[53, 54] in the generalized-gradient approximation[55]. The gamma-centered *k* mesh was taken to be 10x10x10. For selected pressure, the non-spin polarized electronic structures were calculated for the optimized lattice parameters and As position using the WIEN2K[56] implementation of the full potential linearized augmented plane wave method within the PBE generalized gradient approximation.

### III. Results and Discussion

#### A. High-Pressure X-ray Diffraction

Our high-pressure x-ray diffraction measurements up to 40 GPa show the hcT phase transition starting near 5 GPa, as well as indications of a second, theoretically predicted phase transition into the fcT phase near 22 GPa (see figure 1)[20, 22]. At the hcT phase transition there is a broad expansion of the *a*-axis lattice parameter, starting at 5 GPa and increasing until 7.3 GPa, where the lattice begins contracting with pressure again. Based on our data, there is not a distinct contraction of the c-axis at the hcT transition, as has been seen in CaK1144 and predicted by theory, though there is a more rapid decrease in the c-axis lattice parameter with pressure over the same range that we see the a-axis expansion, which may be correlated with the hcT [20, 22]. This expansion of the a-axis is also commensurate with the disappearance of superconductivity that we see in our electrical transport measurements. At the fcT phase transition we again see an expansion of the *a*-axis, though this time it is much more broad (~5 GPa wide) than at the hcT transition. Additionally, we see a slight contraction along the *c*-axis, but it doesn't occur until 24.7 GPa, nearly in the middle of the a-axis expansion region. There is rough agreement between our data and theoretical models at both transitions, though the experimental data shows the expansion of the *a*-axis at the fcT transition much more clearly (see Figs. 1a,b). Comparing to the predicted fcT phase transition in $CaKFe_4As_4$ versus $CaRbFe_4As_4$, there was a predicted fcT transition for the K compound at ~12 GPa, versus the Rb compound transition at ~26 GPa[20, 22]. This increase in the fcT transition may be explained by the larger atomic radius of Rb over that of K (2.48Å vs 2.27Å, respectively) since the transition is driven by the As-As bonding across the Rb/K layer. Indeed, our theoretical calculations of the As-As spacing across the Ca and Rb atoms, shown in figure 2, predict the first collapse across the Ca atom near 6 GPa, but the second collapse across the Rb atom is much less distinct, though there is a decrease near 24 GPa that may correspond with the collapse (see also Fig. S2). The As-As bonding is also evidenced in the electronic structure by the overlapping of the As-$4p_z$ electron density as shown in Fig. S4 of Ref.[20]. This has similar correlations with the room temperature resistivities of the compounds investigated by Iyo et al. [18], although the superconducting transition temperatures are not proportional to atomic radius, which may indicate that the cT transition pressures are influenced by electronic correlation effects as well.

### B. High-Pressure Electrical Transport Measurements

#### 1. Pressure Effects

The $AeA$Fe$_4$As$_4$ ($Ae$=Alkali earth and $A$= Alkali elements) compounds form as alternating $A$ and $Ae$ layers stacked across Fe$_2$As$_2$ nets (Fig. 2, inset)[18]. Ambient pressure electrical transport measurements of CaRbFe$_4$As$_4$ (Fig. 3.) show the superconducting transition temperatures near 35K, which is consistent with all of the $AeA$Fe$_4$As$_4$ compounds. Pressure tuning of the lattice spacing is of particular interest in these compounds since the $A$ and $Ae$ ions should interact differently based on the different spacing between the Fe$_2$As$_2$ layers driven by cation size [18]. High-pressure electrical transport, both longitudinal and transverse, was performed using our eight-probe designer diamond anvil cell to investigate the superconducting state of CaRbFe$_4$As$_4$ as a function of pressure (see Fig. S3 for measurement configuration). Resistance as a function of temperature from 300K to 2K shows the superconducting phase onset temperature ($T_c$) at 35K at 3.1 GPa (Fig. 4a), the same as ambient $T_c$ in CaKFe$_4$As$_4$ and several of the other 1144[19, 22]. The resistance of CaRbFe$_4$As$_4$ drops as a function of increasing pressure, yet with different rates for different pressure regions. The room-temperature resistance decreases linearly as a function of pressure for P≤ 6.2 GPa, then at a decreased rate for 7.6 GPa≤P≤ 20.7 GPa, and finally, is nearly pressure independent for P≥ 22.4 GPa (Fig. 4a and Fig. 5a). These three regions correspond roughly with the phase transitions we observe for the half-collapsed tetragonal (hcT) phase near 6 GPa and the fully-collapsed tetragonal (fcT) phase above 22 GPa, and also aligns well with the measured hcT phase transition reported for CaK1144 by Kaulaurachi et al.[20].

To compare the effects of pressure on the resistivity in these different regions we have plotted the resistivity as a function of pressure at various temperatures above the superconducting transition in figure 4b. The resistivity as a function of pressure clearly shows the hcT and fcT transitions as sharp decreases in resistivity at 6 and 22 GPa, respectively. Though this is born out in Figs. 4 a and b, a decrease in resistivity before the transition is the opposite to previous literature, where there was an increase in the resistivity as they approached the transition[20, 57]. Though we do not have clear understanding of why this is the case, it could, as reference [57] concluded, be due to a difference in pressure homogeneity. The inset of figure 4a shows the superconducting transition for P=3.1-7.6 GPa in a temperature range closer to $T_c$ in which the traces have been offset vertically in order to show the suppression of $T_c$ with pressure. There is also a change in $dT_c/dP$ seen near the same pressures as the resistance transitions (Fig. 5). This can be seen in the $T_c$ versus pressure plot in figure 5c as the slope turns over near 6 GPa, and then by P = 17.2 GPa $T_c$ is suppressed below 2K (error bars for $T_c$ in Fig. 5c are based on the variation due to fitting). The change in $dT_c/dP$ near 6 GPa also correlates well with the structural collapse seen in our x-ray diffraction measurements, and the suppression of $T_c$ is close to the second collapse measured near 22 GPa. Interestingly, we do not see a "V-shaped" phase diagram, as was seen in the *A*122 (A= K, Rb, and Cs) compounds, suggesting that the pairing symmetry may be preserved in CaRb1144[58]. Instead, it seems more reminiscent of the abrupt disappearance of superconductivity seen in $Ca(Fe_{1-x}Co_x)_2As_2$ at the transition into the fcT phase [33, 59].

2. *Magnetic Field Effects*

We also investigated the effects of high magnetic fields on the transport properties of $CaRbFe_4As_4$. A study of the suppression of $T_c$ with increasing magnetic field was seen at all of the pressures for which the superconducting state was measured. $T_c$ decreases at a rate of $dT_c/dH$= -0.3 K/T at P = 3.1 GPa, and follows a similar rate of suppression for all of the pressures studied (Fig. 4c.). As was also seen in Fig. 3, there is a small amount of broadening in the superconducting transition with applied field (see figure S4), which could be due to magnetic fluctuations or vortex effects. If we compare the suppression of $T_c$ as a function of pressure for $CaRbFe_4As_4$ and $CaKFe_4As_4$, we can see that the pair-breaking mechanism has similar field dependences, since both are linearly suppressed in fields up to 15 T, although this may be trending towards a more non-linear dependence at 12.8 and 15.2 GPa (Fig. 4c)[19]. Hall effect measurements were performed in magnetic fields up to 10 T for pressures up to 32.6 GPa (Fig. 4d). The dominant carrier type changes between 4.7 and 7.6 GPa at 50K, as seen in Figure 4d, followed by an increasing Hall resistance up to 17.2 GPa, at which point the resistance begins to decrease with increasing pressure. These changes in carrier type and pressure dependence correspond with the changes that are observed in the longitudinal resistance as well (Fig. 4a). The various pressure dependencies are plotted together in figure 5, to show that many of the phenomena seen in this material take place roughly at the phase boundaries where we see the structural phase transitions (hcT and fcT, marked as blue dashed lines in Fig. 5).

### C. Density Functional Calculations

Calculations were performed at several pressures within the experimental range to investigate the changes to the electronic structure and possible structural transitions. Figure 1a,b demonstrates the agreement of theory and experiment on what pressure the hcT structural collapse occurs, though theory does not capture the fcT transition to the degree we measure experimentally. Both theory and experiment show the hcT transition in the expansion of the a-axis near 6 GPa, though the theoretical calculations predict a larger increase than is seen in the experimental data (Figs. 1a,b). Whereas for the fcT transition, theory predicts only a minor expansion, but the experimental data shows the expansion of the a-axis and is significantly larger than that predicted by the theoretical calculations and also begins near 22 GPa, whereas the theoretical calculations only show a very minor expansion of the *a*-axis, and not until closer to 26 GPa. The contraction of the *c*-axis is much smaller compared to the expansion of the *a*-axis, seen as a broad collapse at the hcT, and is barely noticeable at the fcT, though theory and experiment show good agreement both on the pressure at which the collapses occur, as well as the amount of contraction along the *c*-axis. Changes in the pressure-dependent electronic structure (Fig. S6) are illustrated more clearly in the predicted Fermi surfaces (Fig. 6) and show significant changes progressing through the hcT and fcT phases. The change in the hole pockets at the $\Gamma$ point ($\mathbf{k}$= (0,0)) show the most dramatic change with pressure (Fig. 6a,b), though there is a clear modification to the electron pockets at $\mathbf{M}$. Interestingly, when the system goes through the hcT phase transition at P= 6 GPa only the largest of the hole pockets changes significantly. Only when the system goes through the fcT transition is there a major change to the smaller hole pockets (see Figs. 6c,d), which suggests that the nesting is actually between one of the smaller hole pockets at $\Gamma$ and the larger electron pockets at the $\mathbf{M}$ point ($\mathbf{k}$= ($\pm\pi,\pm\pi$)). This theoretically predicted change in the bandstructure is confirmed by our Hall measurements which show a reversal of the Hall coefficient as a function of pressure near P= 20 GPa (Figs. 4d and 5b). In many of the 122 systems the nesting of these pockets is believed to have a strong influence on the superconducting state of the material.[60-62] Indeed, Mou et al., based on ARPES measurements and density functional theory, have proposed that the degree of nesting of the hole and electron pockets in $CaKFe_4As_4$ determines the maximum value of the superconducting gap[32]. They also found that the nesting in $CaKFe_4As_4$ was between the midsize hole pocket, $\beta$, at $\Gamma$ and the electron pocket, $\delta$, at $\mathbf{M}$. These pockets are similarly sized to the midsized hole pockets at the $\Gamma$ point in $CaRbFe_4As_4$, lending further evidence that the nesting in $CaRbFe_4As_4$ is between those pockets and not with the larger hole pocket with the greater pressure dependence. Given that our electronic structure calculations show a similar Fermi surface to Mou et al., and their agreement with our experimental findings, there is compelling evidence that Fermi surface nesting in the 1144 systems may be a key correlation between superconductivity and the structural collapses[63].

### D. Conclusion

In summary, we have performed high-pressure x-ray diffraction and magnetotransport measurements on $CaRbFe_4As_4$ and observed phase transitions that are consistent with the predicted half- and fully-collapsed tetragonal phases. Our x-ray diffraction measurements show the hcT phase transition near 6 GPa with the broad expansion along the *a*-axis, in good agreement with theoretical predictions, though we see a minimal contraction along the *c*-axis, which was not the case with theory. The fcT transition near 22 GPa is evidenced by another broad expansion of the a-axis, but with an accompanying collapse in the c-axis that is much smaller, as seen in the high-pressure x-ray diffraction data, but was not predicted by our theoretical calculations. This transition agrees much better with the predictions of Borisov et al., though they predict a higher transition pressure than we measured. Magnetotransport measurements show the superconducting phase onset temperature ($T_c$) at 35K at 3.1 GPa. Magnetic-field suppression of $T_c$ shows a rate of $dT_c/dH$=-0.3 K/T, similar to other 1144s, until it is undetectable for P> 17 GPa. Changes in the room-temperature resistance and Hall effect suggest that there are three pressure regimes, P< 6 GPa, 6<P< 20 GPa, P> 20 GPa, with distinct electronic behavior, which correspond well with the tetragonal, half-collapsed tetragonal and fully-collapsed tetragonal phases seen via our x-ray diffraction measurements. Taken together with our density functional calculations, and other experimental and theoretical studies of the 1144 high-pressure phases, we have made the first observation of the fully-collapsed tetragonal phase in a $AeAFe_4As_4$ compound, adding to the growing understanding of this new class of iron-based superconductors.

National Laboratory. HPCAT operations are supported by DOE-NNSA's Office of Experimental Sciences.  The Advanced Photon Source is a U.S. Department of Energy (DOE) Office of Science User Facility operated for the DOE Office of Science by Argonne National Laboratory under Contract No. DE-AC02-06CH11357
 Beamtime was provided by the Carnegie DOE-Alliance Center (CDAC). YKV acknowledges support from DOE-NNSA Grant No. DE-NA0002928. This material is based upon work supported by the U.S. Department of Energy, Office of Science, Office of Workforce Development for Teachers and Scientists, Office of Science Graduate Student Research (SCGSR) program. The SCGSR program is administered by the Oak Ridge Institute for Science and Education for the DOE under contract number DE-SC0014664. Research at the University of Maryland was supported by the Air Force Office of Scientific Research (Award No. FA9550-14-1-0332) and the Gordon and Betty Moore Foundation's EPiQS Initiative through Grant No. GBMF4419.

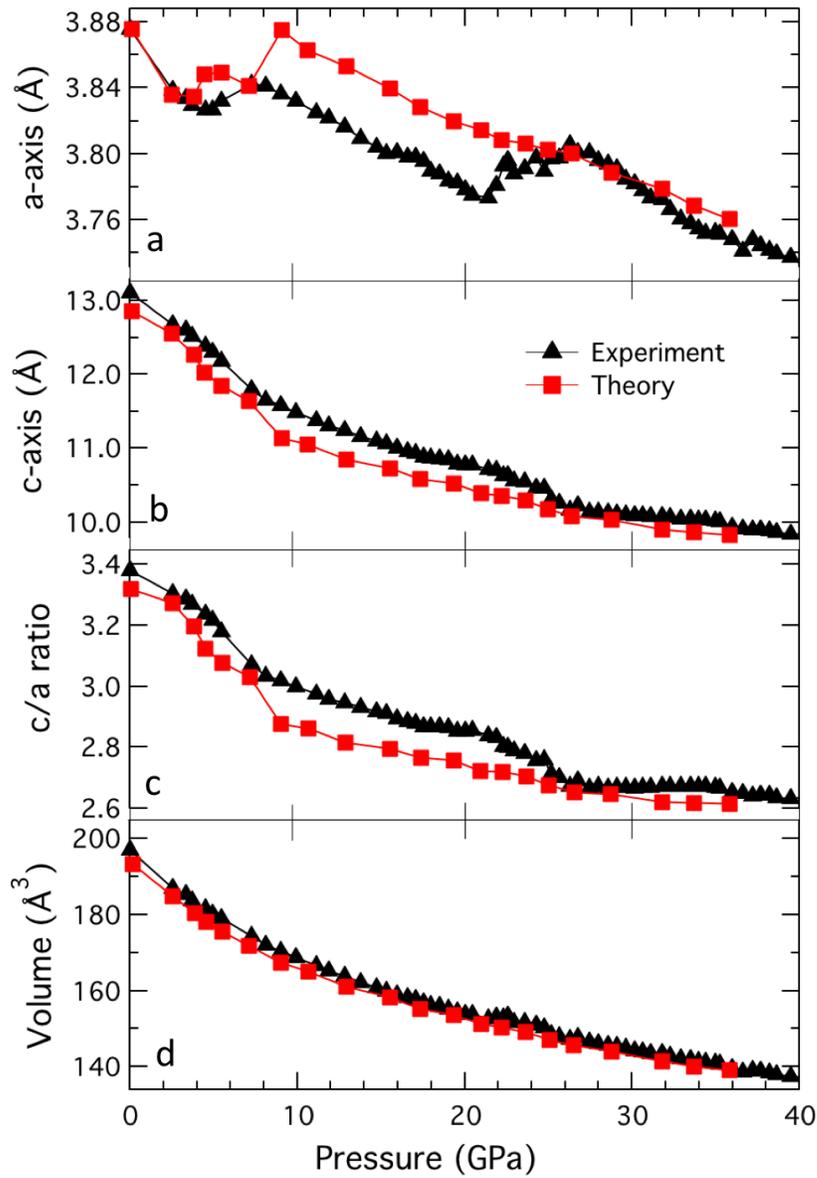

*Figure 1 Pressure dependence of lattice parameters from x-ray diffraction measurements and theoretical calculations. (a) Experimental and theoretical data showing the a-axis lattice expansion near 6 GPa at the half collapse, and the full collapse near 22 GPa. (b) The c-axis lattice parameter shows a much smaller effect, but still at the same pressures that the a-axis changes occur with a much better agreement between experiment and theory. (c) Plot of the c/a ratio highlighting the dominance of the a-axis collapse near 6 GPa. The full collapse near 22 GPa is a much smaller and uniform between the a- and c-axis.*

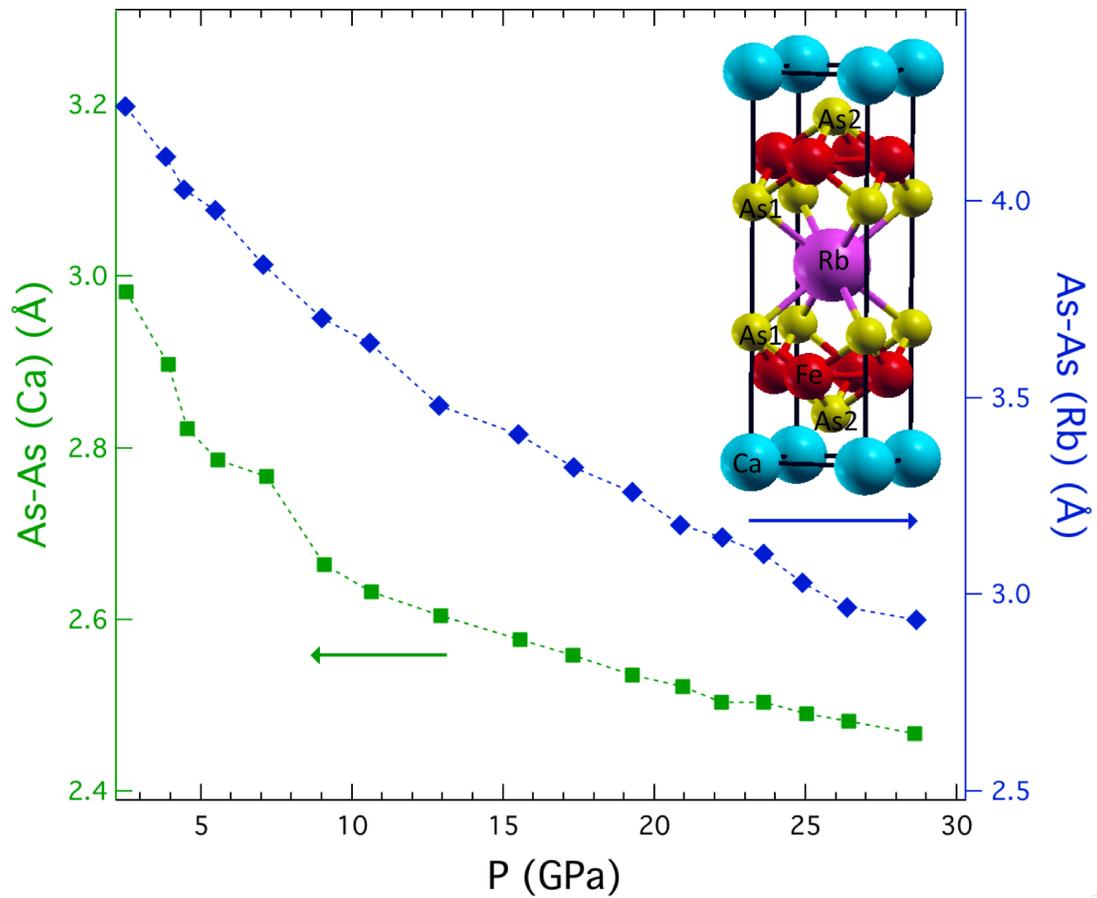

*Figure 2 Theoretical calculations of the spacing between arsenic layers as a function of pressure across the calcium layer (green, left axis) and across the rubidium layer (blue, right axis). Inset shows the unit cell of CaRbFe$_4$As$_4$ where As1 and As2 illustrate the "As-As" distances across the calcium and rubidium atoms.*

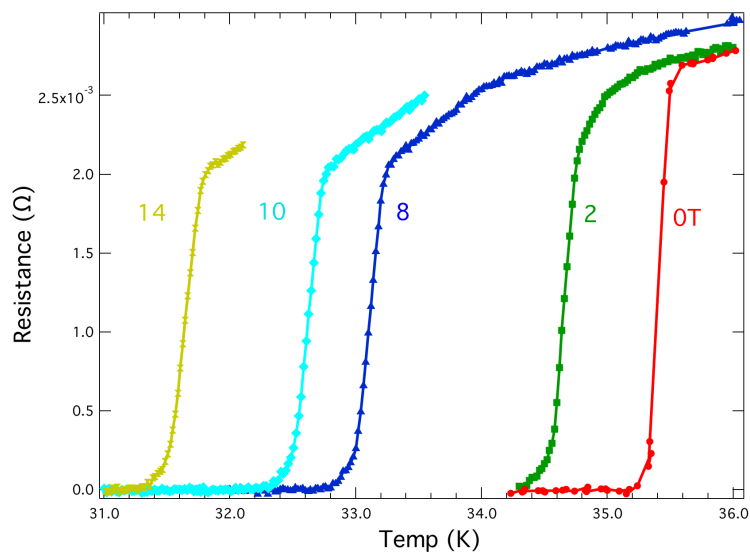

*Figure 3 Ambient-pressure resistance versus temperature showing the suppression of $T_c$ as a function of magnetic field up to 14T.*

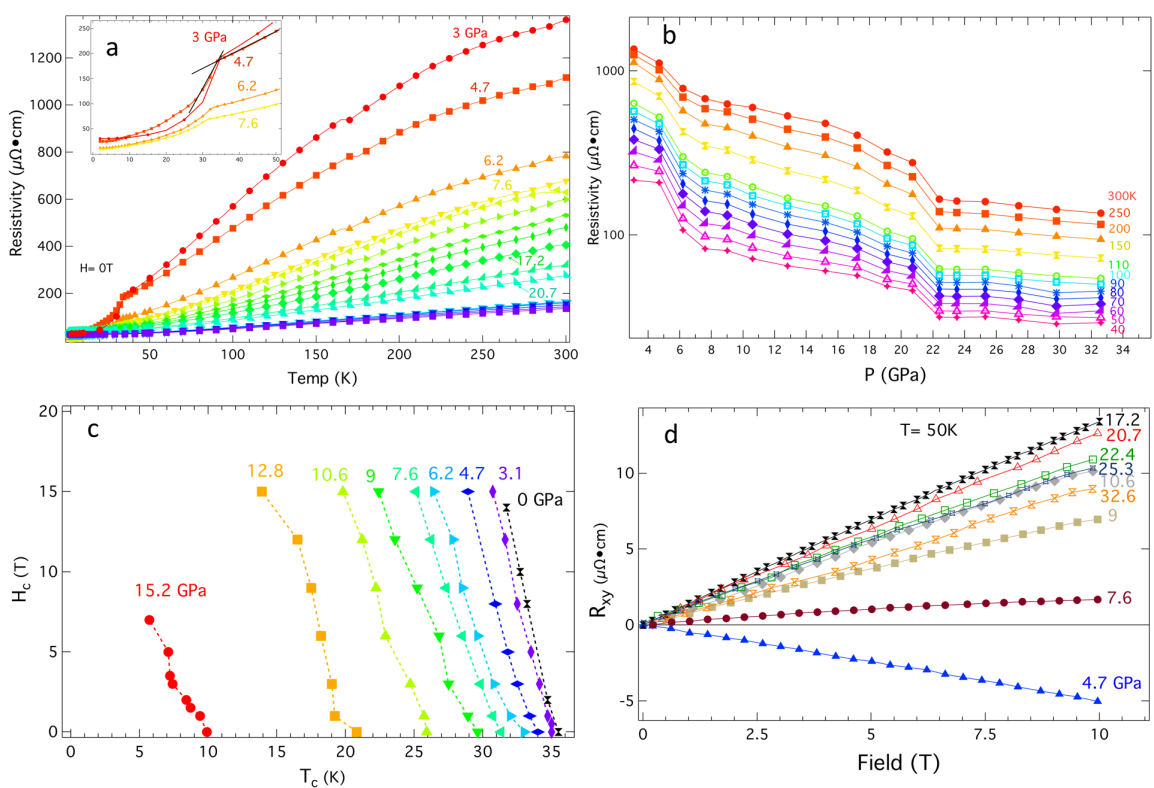

*Figure 4 Effects of pressure, temperature and magnetic fields on $CaRbFe_4As_4$. (a) Longitudinal resistivity ($R_{xx}$) from 300K to 2K as a function of increasing pressure. Jumps in room-temperature resistivity roughly correspond to the half- and fully-collapsed tetragonal phase transitions. (b) Resistance as a function of pressure at selected*

temperatures above the superconducting transition. The hcT and fcT transitions can be seen clearly as the resistance drops near $P_c$. (c) Pressure dependence of the critical field ($H_c$) up to 15.1 GPa (d)Transverse resistance ($R_{xy}$) measurements of the Hall effect at 50K showing a change in dominant carrier type near 6.2 GPa (P=6.2 GPa is at T=2K, but included to show trend) and a change in pressure dependence near 17.2 GPa.

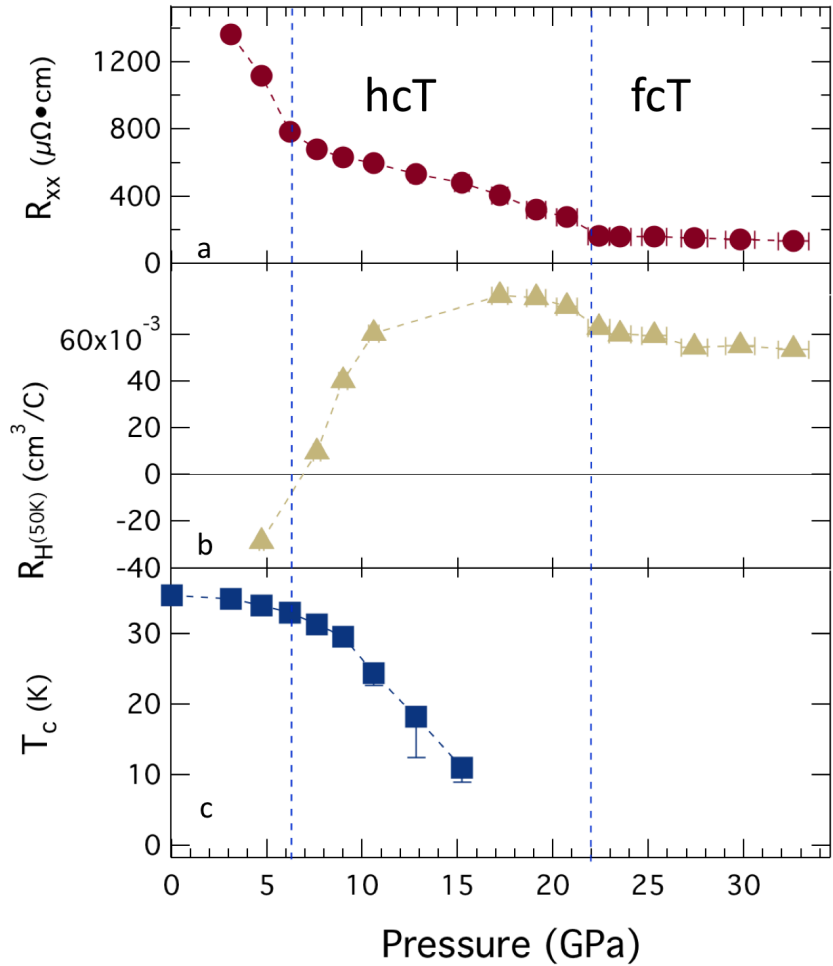

Figure 5 Half-collapsed (hcT) (P~6GPa) and fully-collapsed (fcT) (P~22GPa) phase transitions shown in (a) $R_{xx}$ (300K) (b) $R_H$ and (c) $T_c$ all as a function of pressure. The blue dashed lines indicate where the hcT and fcT transitions are seen in our x-ray diffraction measurements. There is very good agreement between the two measurement techniques, especially considering the small differences could be explained by the use of two different pressure transmitting media. Error bars are described in the text; where not visible, error bars are smaller than the symbols.

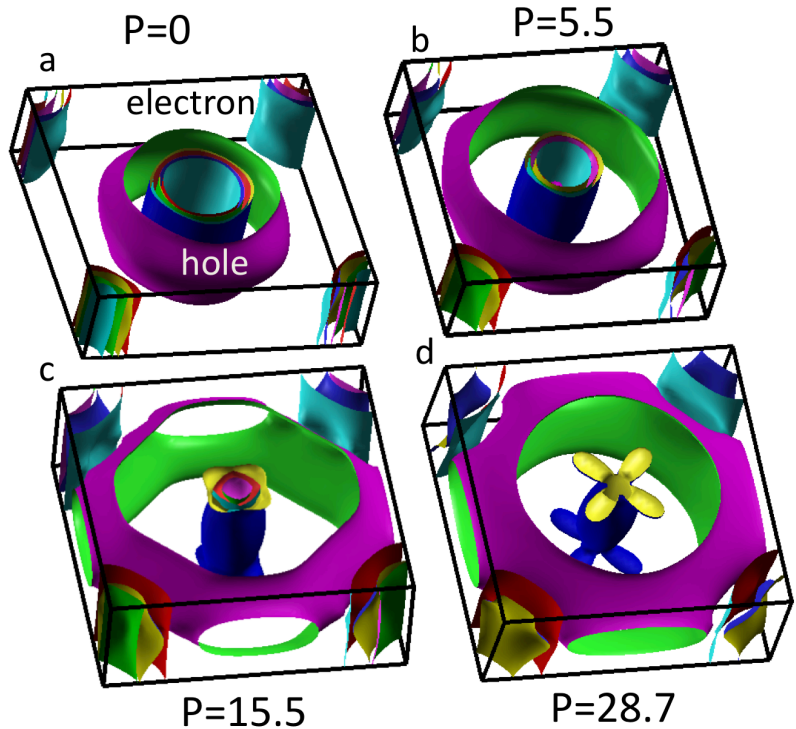

*Figure 6 Fermi surface changes in CaRbFe$_4$As$_4$ as it progresses through the half-and full-collapse. Major reconstruction of the hole pockets seems to be the driving factor based on correlation effects.*